\documentclass{emulateapj} 
\usepackage{apjfonts}
\usepackage{amsmath}
\begin{document}
%\received{}
%\accepted{}
%\revised{}
\slugcomment{ApJ (Letters), in press}
\newcommand{\LeeZinn}{\mathcal{L}}
%%\documentclass[12pt,preprint]{aastex}

%% manuscript produces a one-column, double-spaced document:

%\documentclass[manuscript]{aastex}

%% preprint2 produces a double-column, single-spaced document:

%\documentclass[preprint2]{aastex}

%% Sometimes a paper's abstract is too long to fit on the
%% title page in preprint2 mode. When that is the case,
%% use the longabstract style option.

%% \documentclass[preprint2,longabstract]{aastex}

%% If you want to create your own macros, you can do so
%% using \newcommand. Your macros should appear before
%% the \begin{document} command.
%%
%% If you are submitting to a journal that translates manuscripts
%% into SGML, you need to follow certain guidelines when preparing
%% your macros. See the AASTeX v5.x Author Guide
%% for information.

%\begin{document}

\shortauthors{M. Catelan \& C. Cort\'es} 
\shorttitle{RR Lyr: Evidence for Overluminosity}

\title{Evidence for an Overluminosity of the Variable Star RR Lyr,\\ 
       and a Revised Distance to the LMC}

\author{M. Catelan\altaffilmark{1}}

%\and

\author{C. Cort\'es\altaffilmark{1,2}}

\altaffiltext{1}{Departamento de Astronom\'{i}a y Astrof\'{i}sica, Pontificia Universidad
Cat\'olica de Chile, Santiago, Chile; mcatelan@astro.puc.cl}
%Av. Vicu\~na Mackena 4860, 782-0436 Macul, 

\altaffiltext{2}{Departamento de F\'{i}sica Te\'orica e Experimental, Universidade Federal 
do Rio Grande do Norte, Natal, RN, Brazil; cristian@dfte.ufrn.br}
%Campus Universit\'ario,  59072-970 

\begin{abstract}
We use theoretical models to establish a tight relationship for the absolute 
magnitudes of RR Lyrae stars as a function of their periods and Str\"omgren 
pseudo-color $c_0 \equiv (u\!-\!v)_0 - (v\!-\!b)_0$. 
Applying this to RR Lyr, and comparing the result with 
the predicted average absolute magnitude for stars of similar metallicity 
from the same models, yields an overluminosity of $0.064\pm 0.013$~mag in 
Str\"omgren $y$ (and thus similarly in $V$) for RR Lyr. Based on a revised 
value for RR Lyr's trigonometric parallax, and on a newly derived reddening 
value of $E(\bv) = 0.015\pm 0.020$, we provide a corrected relationship between 
average absolute magnitude and metallicity for RR Lyrae stars that takes 
RR Lyr's evolutionary status fully into account for the first time. Applying 
this relationship to the LMC, we derive a revised true distance modulus 
of $(m-M)_0 = 18.44 \pm 0.11$. 
\end{abstract}


\section{Introduction}
RR Lyrae (RRL) stars are radially pulsating 
variable stars with periods in the $0.2-1.0$~d range, 
and visual amplitudes $\lesssim 2$~mag. 
They are accordingly 
easily identified, and play a key role as the cornerstone of the 
Population~II distance scale. They are extensively used to 
determine distances to old and sufficiently metal-poor systems, 
where they are commonly found in large numbers. In particular,   
RRL are present in globular clusters (GC's) and the dwarf 
galaxies in the neighborhood of the Milky Way \citep[e.g.,][]{cgea07},  
and have also been identified in the M31 field 
\citep[e.g.,][]{tbea04,adea04}, in some M31 companions 
\citep[e.g.,][]{bpea05}, and in at least 4 M31 GC's \citep{gcea01}. 

While period-luminosity (PL) relations for RRL stars in $JHK$ have 
been known and studied for a long time  
\citep*[e.g.,][]{alea86,mcea04,mdpea06}, 
only recently have such relations been extended to bluer filters. 
In particular, \citet{cc07} have  carried out a theoretical 
study of RRL stars in the \citet{bs63} $uvby$ passband system, 
concluding that several relationships involving the (rather 
reddening-independent) {\em pseudo-color} 
$c_0 \equiv (u\!-\!v)_0 - (v\!-\!b)_0$ are present that may prove 
helpful in deriving distances to RRL stars. As well known,  
the Str\"omgren system represents an invaluable tool in the study of 
the physical parameters of stars, including temperature, 
surface gravity, metallicity, and (cluster) ages 
\citep[e.g.,][]{sn89,fgea00}. 

In this {\em Letter} we provide, for the first time, a method to precisely 
estimate the degree of overluminosity in the visual for even field RRL stars. 
This is based on a comparison between the predicted average RRL 
absolute magnitude for the star's metallicity, on the one hand, and its  
own precise absolute magnitude, on the other~-- the latter being 
derived based solely on $c_0$ and the pulsation period.

\section{The Method}
The chemical composition of RR Lyr was studied by \citet{gcea95}, who derived 
${\rm [Fe/H]} = -1.39$, ${\rm [\alpha/Fe]} = +0.31$ for the star. According 
to \citet{abea01}, this [Fe/H] value is in the \citet[][ ZW84]{zw84}
scale; this translates to ${\rm [Fe/H]} = -1.16$ in the \citet[][ CG97]{cg97} 
scale. Based on \citet*{msea93}, we estimate 
a metallicity $Z = 1.4 \times 10^{-3}$ or $2.3 \times 10^{-3}$ in the two 
scales, respectively. Our results will thus be based 
on models computed for a metallicity $Z = 2 \times 10^{-3}$. 

\begin{figure}[t]
  \centerline{
  \includegraphics*[width=3.20in]{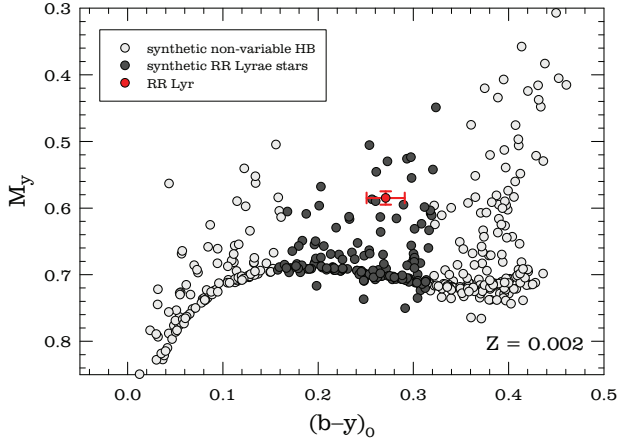}
  }
  \caption{Simulated CMD for an intermediate HB type and a chemical composition
           $Y_{\rm MS} = 0.23$, $Z = 0.002$. Non-variable HB stars are shown 
	   as {\em light gray circles}, whereas RRL variables are indicated 
	   as {\em dark gray circles}. The position of RR Lyr is 
	   shown as a {\em red circle with error bars} (see text). 
   }
      \label{fig:CMD}
\end{figure}

We have computed extensive sets of horizontal branch (HB) simulations 
following the same procedures already described elsewhere \citep{mcea04,cc07}. 
The main sequence helium abundance in 
our models is $Y_{\rm MS} = 0.23$. In Figure~\ref{fig:CMD} we show 
a synthetic $M_y$, $b\!-\!y$ CMD, for a rather even HB type. 
The blue edge of the instability strip (IS) is based on equation~(1) 
in \citet{fcea87}, shifted in temperature by $-200$~K. The IS width 
was been taken as $\Delta\log T_{\rm eff}=0.075$; this defines the IS 
red edge. These choices provide good agreement with recent theoretical 
prescriptions and the observations \citep[see \S6 in][]{mc04}.

RRL-based distances are usually derived using a simple relation 
between {\em average} RRL $V$-band magnitude and metallicity. 
However, RR Lyr is the only RRL-type star with a sufficiently accurate 
trigonometric parallax for distance determinations  
\citep{gbea02}. This star has accordingly been used by several authors to place 
the zero point of the $\langle M_V\rangle - {\rm [Fe/H]}$ relation on a firmer 
footing \citep[e.g.,][]{cc03}. Unfortunately, RR Lyr's evolutionary 
status remains at present unknown, and therefore it is still 
unclear whether it is representative of the underlying stellar 
population to which it is associated. This is an important issue since, 
as can be seen from Figure~\ref{fig:CMD}, a scatter in RRL 
magnitudes (by up to a few tenths of a magnitude; see also \citeauthor{as90}
\citeyear{as90}) is {\em always} expected within any single population, 
and therefore distances based on current calibrations of the 
$\langle M_V\rangle - {\rm [Fe/H]}$ relation may be correspondingly 
in error \citep[e.g.,][]{mc05}. 

\begin{figure}[t]
  \centerline{
  \includegraphics*[width=3.20in]{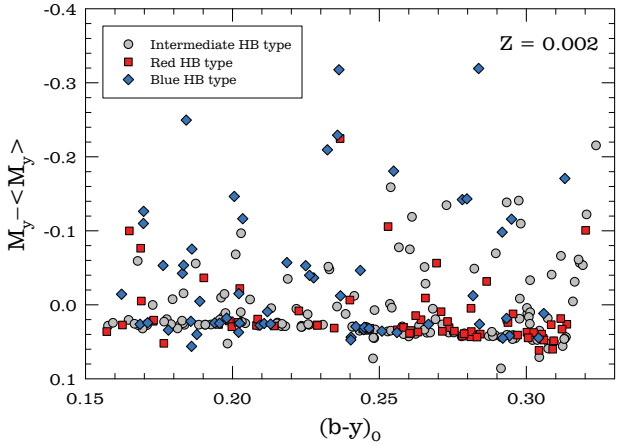}
  } 
  \caption{Difference between each RRL star's individual $M_y$ value and 
           the average over all 8114 synthetic RRL stars in our simulations, for 
	   a blue HB simulation ({\em blue diamond signs}), an even HB simulation 
	   ({\em gray circles}) and a red HB simulation ({\em red squares}). 
   }
      \label{fig:DifCMD}
\end{figure}

To emphasize this point, Figure~\ref{fig:DifCMD} compares the difference in 
absolute magnitude between individual RRL stars in three HB simulations 
with widely different HB morphologies and the average HB absolute magnitude 
computed for a large sample of RRL stars in extensive HB simulations 
covering the full range of observed HB morphologies 
($\langle M_y\rangle = 0.6614 \pm 0.0008$ and median 
$\overline{M_y} = 0.6896$, based on 8114 synthetic stars).

%                                                One column figure
%----------------------------------------------------------- Fig. 1

%                                                One column figure
%----------------------------------------------------------- Fig. 2

Up to now, no method had been available to infer the evolutionary status 
of individual RRL stars to a precision better than 0.1~mag.  
The Str\"omgren system may provide us with a solution to this problem. 
In particular, \citet{cc07} have shown that, by incorporating 
a {\em pseudo-color} term to the PL relation, exceedingly tight 
relations derive, {\em even for the bluer bandpasses}. We have thus used 
the sample of 8114 synthetic RRL stars mentioned above to search for  
high-precision analytical fits that would provide absolute 
magnitudes to a precision of $\sim 0.01$~mag. We thus found a linear 
dependence on $\log P$ and a cubic dependence on the (log of the) 
pseudo-color $c_0$. Our final relations thus read as follows: 

\begin{equation}
M_{\rm band} = a + b \, \ln c_{0} + c \, (\ln c_{0})^2 + d \, (\ln c_{0})^3 + e \, \log P. 
\label{eq:LOG3LIN}
\end{equation}

\noindent The coefficients of the fits for each of the four $uvby$ bands, 
along with their respective errors, are given in Table~\ref{tab:COEF}. 
The correlation coefficient ranges from $r = 0.992$ to 0.998, and the 
standard errors of the estimates from 0.0072 to 0.0078~mag.

%                                                One column figure
%----------------------------------------------------------- Fig. 3
\begin{figure}[t]
  \centerline{
  \includegraphics*[width=3.20in]{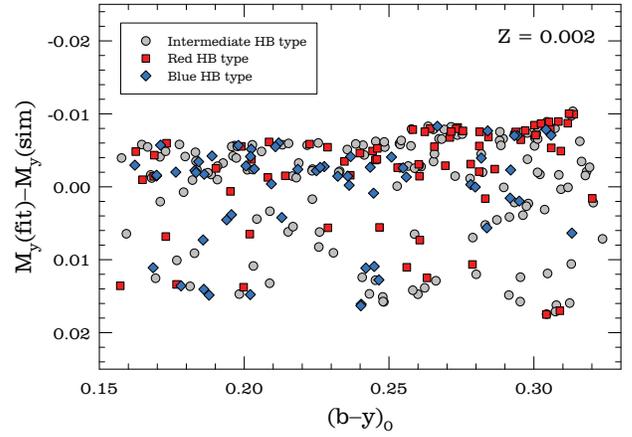}
  }
  \caption{As in Figure~\ref{fig:DifCMD}, but comparing the absolute magnitudes 
       provided by equation~(\ref{eq:LOG3LIN}), here shown as $M_y$(fit), and 
	   the absolute magnitudes of each individual star $M_y$(sim), as provided 
	   by the HB simulations. Note that the scatter is at the level of 
	   $\sigma = 7.2 \times 10^{-3}$~mag only. 
   }
      \label{fig:MyMyEQ}
\end{figure}

\begin{deluxetable}{lccclcc}
\tabletypesize{\footnotesize}
\tablecaption{Coefficients of the Fits}
\tablewidth{0pt}
\tablehead{
\colhead{Coefficient} & \colhead{Value} &  \colhead{Error} &&
\colhead{Coefficient} & \colhead{Value} &  \colhead{Error}}
\startdata
\multicolumn{3}{c}{$u$} && \multicolumn{3}{c}{$v$} \\ 
\cline{1-3}
\cline{5-7}
$a$  & $+1.8000$ & $0.0010$  && $a$ & $+0.4535$ & $0.0009$  \\
$b$  & $-1.1484$ & $0.0029$  && $b$ & $-1.9103$ & $0.0027$  \\
$c$  & $-0.4903$ & $0.0213$  && $c$ & $-0.9250$ & $0.0204$  \\
$d$  & $-1.3840$ & $0.0463$  && $d$ & $-1.4128$ & $0.0443$  \\
$e$  & $-1.7643$ & $0.0026$  && $e$ & $-1.8575$ & $0.0025$  \\
\cline{1-3}
\cline{5-7}
\multicolumn{3}{c}{$b$} && \multicolumn{3}{c}{$y$} \\ 
\cline{1-3}
\cline{5-7}
$a$  & $+0.1070$ & $0.0009$  && $a$ & $-0.1244$  &  0.0009\\
$b$  & $-1.6721$ & $0.0027$  && $b$ & $-1.3546$  &  0.0026\\
$c$  & $-0.8628$ & $0.0197$  && $c$ & $-0.7413$  &  0.0195\\
$d$  & $-1.2972$ & $0.0430$  && $d$ & $-1.3146$  &  0.0425\\
$e$  & $-1.9507$ & $0.0024$  && $e$ & $-2.0577$  &  0.0024
\enddata
\label{tab:COEF}
\end{deluxetable}

To check how well equation~(\ref{eq:LOG3LIN}) is able to reproduce $M_y$
for individual RRL stars, we show, in Figure~\ref{fig:MyMyEQ}, 
the difference between the $M_y$ value implied by this equation and the 
input value from the same HB simulations as in Figure~\ref{fig:DifCMD}. As 
can be clearly seen, the scatter, which in the previous figure amounted to 
several tenths of a magnitude, has been reduced to a few thousandths of a
magnitude only. It is 
thus clear that equation~(\ref{eq:LOG3LIN}) is capable of providing
$M_y$ values for RRL stars with remarkable precision, provided of 
course their periods and $c_0$ values are known. Comparing this with 
$\langle M_y\rangle$ from the same models, one immediately obtains a  
robust estimate of the degree of overluminosity for individual RRL stars.

\section{A New Look at RR Lyr}
\begin{figure}[t]
  \centerline{
  \includegraphics*[width=3.20in]{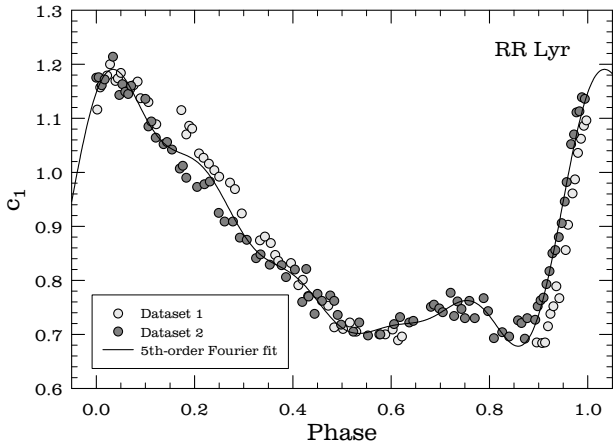}
  }  
  %\plotone{f4.eps}
  \caption{The $c_1$ data for RR Lyr \citep[from][]{ms82} are plotted as a 
           function of the pulsation phase. The two datasets from Siegel are 
	   given with different symbols. A 5th-order Fourier fit to the data 
	   is also shown.
   }
      \label{fig:Fourier}
\end{figure}

Though RR Lyr is an extensively studied star, not many RRL 
surveys have been carried out in the Str\"omgren system to date~-- 
the papers by \citet{ie69} and \citet{ms82} being notable 
exceptions. We shall adopt the latter's dataset in this work. 
\citeauthor{ms82} carried out his $uvby\beta$ study of RR Lyr 
using the 16\,in telescope at Kitt Peak, and provided a $c_1$ light curve. 
This is shown in Figure~\ref{fig:Fourier}, where Siegel's two datasets 
(representing two different phases in the star's $\sim 40$~d-long Blazhko 
cycle) are displayed. 

Naturally, in order to be able to apply our method to 
RR Lyr, we first need to compute a suitable average $c_1$ value for the 
star. In this sense, we have first carried out a Fourier fit to the star's 
observed $c_1$ curve. The result of a 5th-order Fourier decomposition is 
also shown in Figure~\ref{fig:Fourier}. Averaging over this curve in 
magnitude units, we find $(c_1)_{\rm mag} = 0.861 \pm 0.002$. 
Performing the same average in intensity units, we obtain a slightly 
different result, namely $\langle c_1 \rangle = 0.850 \pm 0.002$. 
In like vein, 
$\langle u\rangle -2\langle v\rangle +\langle b\rangle = 0.850\pm 0.010$. 
We have 
performed the same test using Fourier decompositions of different orders, 
with no significant differences in the final average values. Likewise, the 
average $c_1$ value, as derived based on Siegel's (1982) dataset 2 only, 
is not noticeably different from those provided above. (His dataset 1 is 
missing the minimum.) One of the main 
advantages of the present method is indeed the fact that the amplitude 
of the $c_1$ light curve is rather small (0.52~mag, according to the 
5th-order Fourier fit), so that the differences between the several  
procedures to obtain the average quantities appear to be minor. 
We finally adopt for RR Lyr an average $c_1 = 0.855 \pm 0.005$. 

As far as the reddening correction goes, we remind the reader 
that $c_1$ is a rather reddening-insensitive index, with 
$c_0 = c_1 - 0.15\,E(\bv)$. The reddening of RR Lyr is not well 
determined. \citet{asea08} have recently argued that a value
$E(\bv) = 0.02 \pm 0.03$ safely covers all of the available 
determinations. An independent reddening determination is afforded 
by comparing the star's Str\"omgren photometry and our relations 
(see eq.~\ref{eq:LOG3LIN} and Table~1).

%                                                One column figure
%----------------------------------------------------------- Fig. 4

In this sense, we have computed magnitude- and intensity-averaged 
$b-y$, $v-b$, $u-v$, and $u-y$ colors for RR Lyr, and compared 
the values derived using Siegel's (1982) combined datasets and his 
dataset 2 only. We then averaged the results, 
adopting as final average colors for RR Lyr 
the following values: $b\!-\!y = 0.282\pm 0.002$, $u\!-\!v = 1.215\pm 0.003$, 
$v\!-\!b = 0.357\pm 0.002$, and $u\!-\!y = 1.857\pm 0.009$. The error bars 
include the errors of the Fourier fits used to compute the means, as 
well as the dispersion around the mean value given by the different 
definitions of the ``equivalent static star.'' Comparing these values 
with the intrinsic colors implied by our theoretical relations at 
$c_0 = 0.855 \pm 0.005$ and a period $P = 0.5668386$~d 
\citep{kkea06}, using the relationships $E(b\!-\!y)= 0.74\,E(\bv)$, 
$E(u\!-\!v)= 0.65\,E(\bv)$, $E(v\!-\!b)= 0.50\,E(\bv)$, and $E(u\!-\!y)= 1.89\,E(\bv)$
\citep{cm76,jcea04}, and taking a weighted average over the thus derived 
individual $E(\bv)$ values, we find for RR Lyr a revised reddening 
value of 

\begin{equation}
  E(\bv) = 0.015 \pm 0.020, 
\label{eq:REDDS}
\end{equation}

\noindent which also implies a small correction to the adopted $c_0$
value for RR Lyr, which becomes $c_0 = 0.853 \pm 0.006$. 

Inserting these values into equation~(\ref{eq:LOG3LIN}) for $y$, we find 
$M_y = 0.585 \pm 0.010$~mag. Using the average $M_y$ for RRL stars of 
similar metallicity, from \S2, we find that RR Lyr is an overluminous star, 
by about $0.077\pm 0.010$~mag in $y$ (and thus $V$).\footnote{To derive 
an overluminosity in other bandpasses is more tricky, in view of the 
non-zero slope of the HB in planes other than $M_y$, $b-y$. Taking the HB 
slope into due account, we find for RR Lyr overluminosities in $u$, $v$, 
$b$ of $-0.075$, $-0.113$, and $-0.096$~mag, respectively.} The inferred position 
of the star is shown in the CMD of Figure~\ref{fig:CMD}. For an overluminosity 
in $y$ of zero to obtain, we would have needed a $c_0 = 0.801$ for RR Lyr, 
which is many sigma away from the value derived on the basis of Siegel's 
(1982) data. Note that the star's overluminosity, with respect to the 
{\em median}, is even higher: $0.105\pm 0.010$~mag. We emphasize that 
this result is based on a  
differential use of the theoretical models. Within the error bars, 
RR Lyr's overluminosity is also not inconsistent with the recent near-IR 
study by \citet[][ their Fig.~2]{asea08}.

To check the model dependence of this result, we repeated our procedure 
using the \citet{apea04} tracks, corrected as  in \citet{scea04}. 
Based on the resulting HB simulations  
we rederive the parameters of equation~(\ref{eq:LOG3LIN}), leading 
to a relation with $r = 0.993$ and standard error of the estimate 0.006~mag. 
We infer the $M_y$ value for RR Lyr therefrom ($M_y = 0.580$~mag), and 
compare with the value $\langle M_y\rangle = 0.6576$~mag from the same 
simulations. This gives for RR Lyr an overluminosity of 0.077~mag, in 
perfect agreement with the value derived earlier. Using the \citet{ld90} 
models for $Y_{\rm MS} = 0.23$, the overluminosity we find is 0.064~mag, 
again consistent with our previous results. 

\citet{cc07} caution that a star's exact overluminosity will depend on 
the adopted metallicity scale. To safely accomodate the resulting range 
of possible values for RR Lyr, we finally adopt for this star an 
overluminosity in $y$ (and thus $V$) of $0.064\pm 0.013$~mag.

\section{A Revised RRL $\langle M_V \rangle - {\rm [Fe/H]}$ Relation} 
Let us assume, as often done, a relation of the form 
$\langle M_V \rangle = \alpha {\rm [Fe/H]} + \beta$. Here we shall adopt 
a slope $\alpha = 0.23 \pm 0.04$, as suggested in several 
recent reviews of the subject \citep[e.g.,][]{cc03}. What about the 
zero point $\beta$?

\citet{gbea02} obtained, using the {\em Hubble Space Telescope} (HST),  
an accurate value for  RR Lyr's trigonometric parallax, namely 
$\pi_{\rm abs}^{\rm HST} = 3.82\pm 0.20$~mas. Very recently, \citet{vl07} revised 
the trigonometric parallaxes provided by Hipparcos, and arrived at a value 
$\pi_{\rm abs}^{\rm Hip} = 3.46\pm 0.64$~mas for RR Lyr. 
A weighted average of ground-based studies \citep*{valh95} indicates 
a parallax $\pi_{\rm abs}^{\rm ground} = 3.0\pm 1.9$~mas for this star. 
Taking a weighted average of these results, we obtain a 
final value of $\pi_{\rm abs} = 3.78\pm 0.19$~mas for RR Lyr. This implies 
a revised distance modulus of $(m\!-\!M)_0 = 7.11\pm 0.11$~mag for the star. 

In recent work, an intensity-weighted 
mean magnitude of $\langle V \rangle = 7.76$~mag \citep[][ F98]{fea98}
has been adopted for RR Lyr. However, we note that 
this value is based on {\em Hipparcos} photometry, which may require a 
non-trivial transformation to the standard system. For comparison, 
\citet{l94} determines a $\langle V \rangle = 7.66$~mag, and \citet{lea96} 
find instead $\langle V \rangle = 7.74$~mag. \citet[][ GP98]{gp98} 
argue strongly in favor of the {\em Hipparcos}-based magnitudes of 
F98, but propose a (small) reddening-related 
correction: $V_{\rm GP98} = V_{\rm F98} - 0.2 \, E(\bv)$. Using our 
derived reddening value (eq.~\ref{eq:REDDS}), and 
adopting a standard extinction law with $A_V = 3.1 \, E(\bv)$, one finds 
$A_V^{\rm RRL} = 0.045\pm 0.062$~mag. The final, extinction-corrected 
RR Lyr mean magnitude is accordingly 
$\langle V_0 \rangle = 7.710\pm 0.062$~mag. 

As well known, 
the intensity- or magnitude-weighted mean magnitude does not necessarily 
correspond to the magnitude of the ``equivalent static star'' (i.e., the 
magnitude 
the star would have if it were not pulsating): an amplitude-dependent 
correction has to be applied. Such a correction has been obtained by 
\citet*{bea95} on the basis of detailed hydrodynamical models. 
Inspection of the light curves for RR Lyr 
\citep[e.g.,][]{sk00,sea03,kkea06} 
suggests that the amplitude in 
$V$ oscillates in the $0.5-1.1$~mag range. According to 
Table~2 in \citeauthor{bea95}, this is precisely the amplitude range  
over which the intensity-weighted mean magnitude provides the most 
accurate description of the magnitude of the equivalente static star. 
Taking, accordingly, a value $\langle V_0 \rangle = 7.710\pm 0.062$~mag
and a distance modulus $(m\!-\!M)_0 = 7.11\pm 0.11$~mag, one finds an 
absolute magnitude for the star of $M_V = 0.600 \pm 0.126$~mag. 

This, as previously argued, should {\em not} be viewed as representative 
of the average absolute magnitude of RRL stars of similar metallicity: 
according to our results, the latter are fainter (on average) than RRL 
by about $0.064\pm 0.013$~mag. Therefore, the average absolute magnitude 
of RRL stars of metallicity similar to RR Lyr's should be 

\begin{equation}
\langle M_V \rangle = 0.664 \pm 0.127 \,\, {\rm mag}. \tag{3}
\label{eq:MVAVE}
\end{equation}

Adopting for RR Lyr the metallicity values provided in \S2, one then 
finds the following final relations: 

\begin{align}
\langle M_V \rangle & = & (0.23 \pm 0.04) \, {\rm [Fe/H]_{ZW84}} + (0.984 \pm 0.127), \tag{4a} %\eqnum{4a}
\label{eq:ZW} \\ 
\langle M_V \rangle & = & (0.23 \pm 0.04) \, {\rm [Fe/H]_{CG97}} + (0.931 \pm 0.127). \tag{4b} %\eqnum{4b}
\label{eq:CG} 
\end{align}

To close, we note, in passing, that the result given in equation~(\ref{eq:MVAVE}) 
is in reasonable agreement with the average $M_y$ value at $Z = 0.002$ 
derived from our HB simulations (\S2).

\section{A Revised LMC Distance Modulus} 
Equation~(\ref{eq:ZW}) implies an $M_V = 0.644\pm 0.141$~mag at 
${\rm [Fe/H]} = -1.48\pm 0.07$, which is the 
mean metallicity derived for LMC RRL variables by 
\citet[][ ZW84 scale]{rgea04}. Using a value 
$\langle V_0 \rangle = 19.068\pm 0.102$~mag from \citeauthor*{rgea04}, 
one then finds a true distance modulus for the LMC of 
$(m\!-\!M)_0^{\rm LMC} = 18.42\pm 0.17$. 
If one uses instead the average values for LMC RRL stars independently 
determined by \citet{bea04}, namely ${\rm [Fe/H]} = -1.46\pm 0.09$~dex and 
$\langle V \rangle = 19.45\pm 0.04$~mag, with their favored reddening of 
$E(\bv) = 0.11$~mag, one finds $\langle V_0 \rangle = 19.11\pm 0.04$~mag 
for a $M_V = 0.648\pm 0.142$~mag; thus
$(m\!-\!M)_0^{\rm LMC} = 18.46\pm 0.15$. 
Taking a weighted average over these two results, we arrive at 
the following distance modulus for the LMC:

\begin{equation}
   (m\!-\!M)_0^{\rm LMC} = 18.44\pm 0.11.  \tag{5}
   \label{eq:LMC3}
\end{equation}

%\noindent which is in excellent agreement with the newly derived value 
%from a reanalysis of the Hipparcos parallaxes for Galactic Cepheids 
%\citep{vlea07}. 

\section{Conclusions}
In this {\em Letter}, we have established, based on Str\"omgren photometry, 
that the star RR Lyr is approximately 0.064~mag brighter than the average 
for other RRL stars of similar metallicity. We also derive a new reddening 
value for RR Lyr, namely, $E(\bv) = 0.015 \pm 0.020$. These results, along 
with a newly derived trigonometric parallax for RR Lyr, lead to a revision 
in the RRL absolute magnitude-metallicity relation, and thus to a revised 
distance modulus for the LMC.

\acknowledgments Useful comments by an anonymous referee are gratefully 
acknowledged. We also thank F. Grundahl and H. A. Smith for their comments 
on a draft of this paper. This work is supported by Proyecto FONDECYT 
Regular \#1071002.


\begin{thebibliography}{}
\bibitem[Benedict et al.(2002)]{gbea02}
  Benedict, G. F., et al. 2002, \aj, 123, 473 

%\bibitem[Blazhko(1907)]{b07}
%  Blazhko, S. 1907, Astron. Nachr., 175, 325

\bibitem[Bono et al.(1995)Bono, Caputo, \& Stellingwerf]{bea95}
  Bono, G., Caputo, F., \& Stellingwerf, R. F. 1995, \apjs, 99, 263

\bibitem[Borissova et al.(2004)]{bea04}
  Borissova, J., Minniti, D., Rejkuba, M., Alves, D., Cook, K. H., \& Freeman, 
    K. C. 2004, \aap, 423, 97

\bibitem[Bragaglia et al.(2001)]{abea01}
  Bragaglia, A., Gratton, R. G., Carretta, E., Clementini, G., Di Fabrizio, L., 
    \& Marconi, M. 2001, \aj, 122, 207

\bibitem[Brown et al.(2004)]{tbea04}
  Brown, T. M., Ferguson, H., Smith, E., Kimble, R. A., Sweigart, A. V., Renzini, 
    A., \& Rich, R. M. 2004, \aj, 127, 2738

\bibitem[Cacciari \& Clementini(2003)]{cc03}
  Cacciari, C., \& Clementini, G. 2003, in Stellar Candles for the Extragalactic 
    Distance Scale, ed. D. Alloin \& W. Gieren (Berlin: Springer), 105

%\bibitem[C\'aceres \& Catelan(2007)]{clc07}
%  C\'aceres, C., \& Catelan, M. 2007, \apjs, submitted 

\bibitem[Caputo et al.(1987)]{fcea87} 
  Caputo, F., de Stefanis, P., Paez, E., \& Quarta, M.~L.\ 1987, \aaps, 68, 119 

\bibitem[Carretta \& Gratton(1997)]{cg97}
  Carretta, E., \& Gratton, R. 1997, \aaps, 121, 95 (CG97)

\bibitem[Cassisi et al.(2004)]{scea04}
  Cassisi, S., Castellani, M., Caputo, F., \& Castellani, V. 2004, \aap, 426, 641  
  
\bibitem[Catelan(2004)]{mc04} 
  Catelan, M.\ 2004, \apj, 600, 409 

\bibitem[Catelan(2005)]{mc05}
  Catelan, M. 2005, preprint (astro-ph/0507464) 

\bibitem[Catelan et al.(2004)Catelan, Pritzl, \& Smith]{mcea04} 
  Catelan, M., Pritzl, B.~J., \& Smith, H.~A.\ 2004, \apjs, 154, 633 

%\bibitem[Chaboyer(1999)]{bc99}
%  Chaboyer, B. 1999, in Post-Hipparcos Cosmic Candles, ed. A. Heck \& F. Caputo
%   (Dordrecht: Kluwer), 111 

\bibitem[Clem et al.(2004)]{jcea04} 
  Clem, J.~L., VandenBerg, D.~A., Grundahl, F., \& Bell, R.~A.\ 2004, \aj, 
    127, 1227 

\bibitem[Clementini et al.(1995)]{gcea95}
  Clementini, G., Carretta, E., Gratton, R., Merighi, R., Mould, J. R., \& 
  McCarthy, J. K. 1995, \aj, 110, 2319 

\bibitem[Clementini et al.(2001)]{gcea01}
  Clementini, G., Federici, L., Corsi, C., Cacciari, C., Bellazzini, M., \& 
    Smith, H. A. 2001, \apjl, 559, L109

\bibitem[Cort\'es \& Catelan(2007)]{cc07}
  Cort\'es, C., \& Catelan, M. 2007, \apjs, submitted 

\bibitem[Crawford \& Mandwewala(1976)]{cm76} 
  Crawford, D.~L., \& Mandwewala, N.\ 1976, \pasp, 88, 917 

\bibitem[Del Principe et al.(2006)]{mdpea06} 
  Del Principe, M., et al.\ 2006, \apj, 652, 362 

%\bibitem[Detre \& Szeidl(1973)]{ds73}
%  Detre, L., \& Szeidl, B. 1973, IBVS, 764, 1

\bibitem[Dolphin et al.(2004)]{adea04}
  Dolphin, A. E., Saha, A., Olszewski, E. W., Thim, F., Skillman, E. D., 
    Gallagher, J. S., \& Hoessel, J. 2004, \aj, 127, 875

\bibitem[Epstein(1969)]{ie69}
  Epstein, I. 1969, \aj, 74, 1131

\bibitem[Fernley et al.(1998)]{fea98}
  Fernley, J., Barnes, T. G., Skillen, I., Hawley, S. L., Hanley, C. J., Evans, 
    D. W., Solano, E., \& Garrido, R. 1998, \aap, 330, 515 (F98)

\bibitem[Gould \& Popowski(1998)]{gp98}
  Gould, A., \& Popowski, P. 1998, \apj, 508, 844 (GP98)

\bibitem[Gratton et al.(2004)]{rgea04}
  Gratton, R. G., Bragaglia, A., Clementini, G., Carretta, E., Di Fabrizio, L., 
    Maio, M., \& Taribello, E. 2004, \aap, 421, 937

\bibitem[Greco et al.(2007)]{cgea07}
  Greco, C., et al. 2007, \apj, 670, 332

%\bibitem[Greco et al.(2008)]{cgea08}
%  Greco, C., et al. 2008, \apjl, in press (astro-ph/0712.2241) 

\bibitem[Grundahl et al.(2000)]{fgea00} 
  Grundahl, F., VandenBerg, D.~A., Bell, R.~A., Andersen, M.~I., \& Stetson, P.~B.\ 2000, 
    \aj, 120, 1884 

%\bibitem[Jones et al.(1987)]{rjea87}
%  Jones, R. V., Carney, B. W., Latham, D. W., \& Kurucz, R. L. 1987, \apj, 
%    312, 254

\bibitem[Kolenberg et al.(2006)]{kkea06}
  Kolenberg, K., et al. 2006, \aap, 459, 577

\bibitem[Layden(1994)]{l94}
  Layden, A. C. 1994, \aj, 108, 1016 

\bibitem[Layden et al.(1996)]{lea96}
  Layden, A. C., Hanson, R. B., Hawley, S. L., Klemola, A. R., \& Hanley, C. J. 
    1996, \aj, 112, 2110

\bibitem[Lee \& Demarque(1990)]{ld90}	
  Lee, Y.-W., \& Demarque, P. 1990, \apjs, 73, 709	
	
\bibitem[Longmore et al.(1986)Longmore, Fernley, \& Jameson]{alea86} 
  Longmore, A.~J., Fernley, J.~A., \& Jameson, R.~F.\ 1986, \mnras, 220, 279 

%\bibitem[Perryman et al.(1997)]{pea97}
%  Perryman, M. A. C., et al. 1997, \aap, 323, L49 

\bibitem[Pietrinferni et al.(2004)]{apea04}
  Pietrinferni, A., Cassisi, S., Salaris, M., \& Castelli, F. 2004, \apj, 612, 168

\bibitem[Pritzl et al.(2005)]{bpea05}
  Pritzl, B. J., Armandroff, T. E., Jacoby, G. H., \& Da Costa, G. S. 2005, \aj, 
    129, 2232

\bibitem[Salaris et al.(1993)Salaris, Chieffi, \& Straniero]{msea93} 
  Salaris, M., Chieffi, A., \& Straniero, O.\ 1993, \apj, 414, 580 

\bibitem[Sandage(1990)]{as90}
  Sandage, A. 1990, \apj, 350, 603

\bibitem[Schuster \& Nissen(1989)]{sn89} 
  Schuster, W.~J., \& Nissen, P.~E.\ 1989, \aap, 221, 65 

\bibitem[Siegel(1982)]{ms82}
  Siegel, M. J. 1982, \pasp, 94, 1225

%\bibitem[Smith(1995)]{hs95}
%  Smith, H. A. 1995, RR Lyrae Stars (Cambridge: Cambridge University Press) 

\bibitem[Smith et al.(2003)]{sea03}
  Smith, H. A., et al. 2003, \pasp, 115, 43

\bibitem[Sollima et al.(2008)]{asea08}
  Sollima, A., Cacciari, C., Arkharov, A. A., Larionov, V. M., Gorshanov, D. L., 
    Efimova, N. V., \& Piersimoni, A. 2008, \mnras, in press %(astro-ph/0712.0578)
	
\bibitem[Str{\"o}mgren(1963)]{bs63} 
  Str{\"o}mgren, B.\ 1963, \qjras, 4, 8 

\bibitem[Szeidl \& Koll\'ath(2000)]{sk00}
  Szeidl, B., \& Koll\'ath, Z. 2000, in The Impact of Large-Scale Surveys on 
    Pulsating Star Research, ASP Conf. Ser. 203, ed. L. Szabados \& D. W. Kurtz
    (San Francisco: ASP), 281 

\bibitem[van Altena et al.(1995)van Altena, Lee, \& Hoffleit]{valh95}
  van Altena, W. F., Lee, J. T., \& Hoffleit, E. D. 1995, Yale Parallax Catalog 
   (4th ed.; New Haven: Yale Univ. Obs.) 

\bibitem[van Leeuwen(2007)]{vl07}
  van Leeuwen, F. 2007, Hipparcos, the New Reduction of the Raw Data (Dordrecht: 
    Springer)

%\bibitem[van Leeuwen et al.(2007)]{vlea07}
%  van Leeuwen, F., Feast, M. W., Whitelock, P. A., \& Laney, C. D. 2007, \mnras, 
%    379, 273

\bibitem[Zinn \& West(1984)]{zw84}
  Zinn, R., \& West, M. J. 1984, \apjs, 55, 45 (ZW84)

\end{thebibliography}
\end{document}